\documentclass[aps,prl,floatfix,superscriptaddress,twocolumn,footinbib]{revtex4-2}

\usepackage{amssymb}
\usepackage{amsmath}
\usepackage{amsfonts}
\usepackage{appendix}
\usepackage{bm}
\usepackage{braket}
\usepackage{graphicx}
\usepackage{epsfig}
\usepackage{epstopdf}
\usepackage{balance}
\usepackage{float}
\usepackage[dvipsnames]{xcolor}
\usepackage{calc}
\usepackage{natbib}
\usepackage[colorlinks,
            linkcolor=blue,
            anchorcolor=blue,
            citecolor=blue,
            urlcolor=blue]{hyperref}
\usepackage{lipsum}
\usepackage{enumerate}
\usepackage{enumitem}

\begin{document}
\title{Theory of Integer Quantum Hall Effect in Irrational Magnetic Field}
\author{Zhao-Wen Miao}
	\affiliation{School of Physics and Wuhan National High Magnetic Field Center, Huazhong University of Science and Technology, Wuhan, Hubei 430074, China}
 \author{Chen Zhao}
	\affiliation{School of Physics and Wuhan National High Magnetic Field Center, Huazhong University of Science and Technology, Wuhan, Hubei 430074, China}
 \author{Jin-Hua Gao}
 \email{jinhua@hust.edu.cn}
	\affiliation{School of Physics and Wuhan National High Magnetic Field Center, Huazhong University of Science and Technology, Wuhan, Hubei 430074, China}
 \author{X. C. Xie}
 \affiliation{Interdisciplinary Center for Theoretical Physics and Information Sciences, Fudan University, Shanghai 200433, China}
  \affiliation{International Center for Quantum Materials, School of Physics, Peking University, Beijing 100871, China}
  \affiliation{Hefei National Laboratory, Hefei 230088, China}
  
\begin{abstract} 
 The conventional theory of the integer quantum Hall effect (IQHE) fails for irrational magnetic fields owing to the breakdown of magnetic translational symmetry. Here,  based on the recently proposed incommensurate energy band (IEB) theory, we present a universal IQHE theory that does not rely on magnetic translation symmetry and is applicable to both rational and irrational magnetic fluxes. Using the square lattice as a paradigmatic example, we first show  that the IEB framework provides a superior description of its energy spectrum in a magnetic field, as it explicitly reveals the momentum-space distribution of eigenstates.  Key to our IQHE theory is that each gap in the IEB spectrum is intrinsically labeled by an integer pair \((m,g)\), defined by the corresponding Bragg planes. When the Fermi energy lies within such a gap, the occupied electron states $N_{\text{occ}}$ is determined by the k-space volume enclosed by these Bragg planes, leading to the fundamental relation \(N_{\text{occ}}/N_0 = m(\phi/\phi_0) + g\). Through  St\v{r}eda formula, this leads directly to the quantized Hall conductance \(\sigma_{xy} = m e^2/h\) under arbitrary magnetic fields. Our work resolves the long-standing problem of IQHE under irrational flux, and establishes a new paradigm for IQHE. 
\end{abstract}

\maketitle

\emph{Introduction}---Since von Klitzing's experimental discovery of quantized Hall conductance\cite{QHE1_Klitzing1980}, the study of the integer quantum Hall effect (IQHE) has become one of the core areas of modern condensed matter physics\cite{Avron2003,Prange1987_IQHE2025cite4,Book1_Stone,janssen1994introduction_IQHE2025cite7,Book2_Thouless,Sarma2008_IQHE2025cite5,yoshioka2013quantum_IQHE2025cite6,tong2016lectures_IQHE2025cite8,witten2016three_IQHE2025cite10,von2017quantum_IQHE2025cite9}. Theoretically, the IQHE is profoundly significant because it was the first phenomenon to reveal that the quantum Hall conductance actually originates from the topological (geometrical) properties of the energy bands\cite{TKNN1982,Avron1983_IQHE2025cite12,Q.Niu1985_IQHE2025cite13,Thouless1981_IQHE2025cite14,P.Streda1982_IQHE2025cite15}.

The conventional IQHE theory is based on the concepts of Chern number and magnetic subbands\cite{TKNN1982}. The square lattice in a magnetic field serves as an ideal pedagogical model, which exhibits a unique fractal energy spectrum---the Hofstadter spectrum---due to the competition between the magnetic length and the lattice period $a_0$\cite{Hofstadter1976,Wannier1978,SquareLattice_MacDonald,SquareLattice_PhysRevB.42.8282,Indubala_BOOK}. Moreover, the renowned TKNN formula states that at rational magnetic flux $\phi/\phi_0=p/q$ (per unit cell, $\phi_0$ is the flux quantum, $p$ and $q$ are a pair of coprime integers), the energy band of the square lattice splits into $q$ magnetic subbands, and the Hall conductance of each magnetic subband corresponds precisely to the integral of its Berry curvature, namely the Chern number. Consequently, when the Fermi level lies within an energy gap, the measured Hall conductance corresponds to the sum of the Chern numbers of all the magnetic subbands below the Fermi level.

Despite the tremendous success of the IQHE theory, the IQHE under irrational magnetic flux has remained poorly understood. The fundamental difficulty lies in the fact that when the flux is irrational, the lattice possesses no magnetic translation symmetry. Therefore, Bloch's theorem cannot be applied to describe its electronic states, and the concepts of magnetic subbands and Chern numbers break down. Although some understanding has been achieved using methods such as rational approximation (e.g., continued fractions) or noncommutative geometry\cite{Bellissard1994,Book3_Madore1999}, a universal  theory of the IQHE under irrational flux is still lacking.

The IQHE under  irrational magnetic flux is inherently equivalent to a quasiperiodic lattice problem. Taking the square lattice as an example, in the Landau gauge, its Hamiltonian can be mapped to a one-dimensional atomic chain with an incommensurate periodic potential, known as the Aubry-André-Harper (AAH) model\cite{P.G.Harper1955,AAH1980}. Recently, an incommensurate energy band (IEB) theory has been proposed for the quasiperiodic lattice systems, where  we have successfully generalized the concept of energy bands to quasiperiodic systems without relying on translation symmetry\cite{ZheHe2024,XinyuGuo2024,JinrongChen2025}. The key insight of the IEB theory lies in the fact that for extended states (localized states in momentum space), although the crystal momentum is no longer a good quantum number, we can still use it as an index for the eigenstates of the quasiperiodic lattices. Interestingly, this IEB theory in fact provides a novel perspective for understanding the IQHE under irrational magnetic fields.

\begin{figure*}[tbp]
	\centering
	\includegraphics[width=1\textwidth]{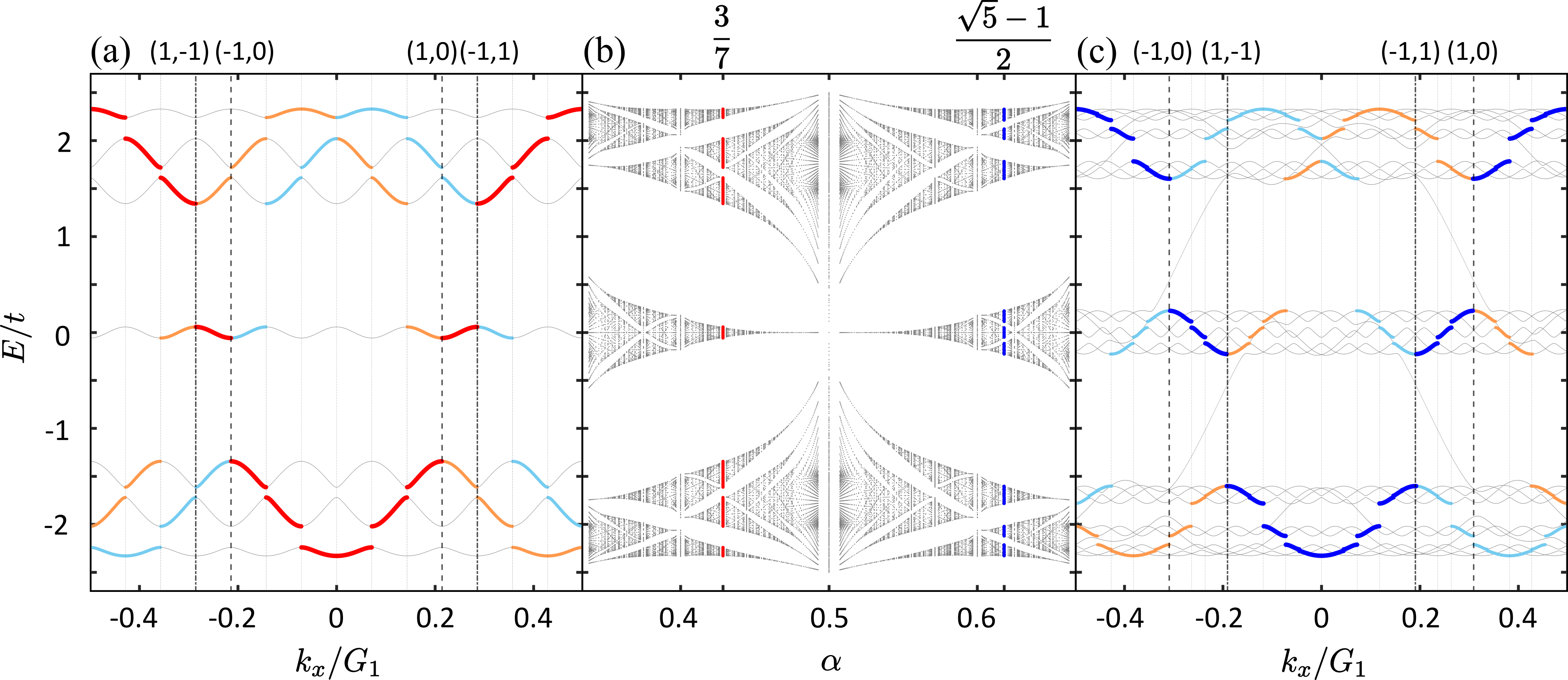}
 \caption{ IEBs and Hofstadter spectrum of a square lattice with  $t_2 = 0.75t_1$. (a) is the IEB of $\alpha = 3/7$ (red lines, rational case) along $k_y=k_x$ direction in BZ.  (b) is the Hofstadter spectrum in selected energy region. (c) is the IEB of $\alpha = (\sqrt{5}-1)/2$  (blue lines, irrational case). In (a) and (c), the replica bands corresponding to $|K_1\rangle$ and $|K_{-1}\rangle$ are highlighted with orange and azure lines, respectively. The vertical dashed lines represent the Bragg planes. Here, the cutoff $n_c = 12$ is used, and  $t_1=t$ as energy unit. }
 \label{Hofstadter1}
\end{figure*}

In this work, we resolve this long-standing theoretical challenge. Based on the IEB theory, we present a unified theory of the IQHE that can describe both rational and irrational magnetic flux cases within a single framework, without invoking the conventional concepts of Chern number and magnetic subbands. Here, we focus on the square lattice as an illustrative example.  The key points of our IQHE theory are summarized as follows:

\begin{enumerate}
    \item The IEB description is significantly superior to the conventional Hofstadter spectrum in characterizing the energy spectrum of a square lattice in a magnetic field, as it provides deeper insights into the eigenstate distribution in momentum space, analogous to conventional energy bands. 

    \item Each energy gap in the IEB spectrum is intrinsically labeled by a pair of integers $(m, g)$. In momentum space, the position of the IEB gaps is determined by the degeneracy points of the so-called replica bands (i.e., Bragg planes), which depend solely on the two integers $(m, g)$ given in the replica band definition.

   \item When the Fermi energy $E_F$ lies within a gap, the total number of the occupied electron  states $N_{occ}$  is determined by the k-space volume enclosed by the corresponding Bragg planes. This yields the fundamental relationship:  $N_{occ}/N_0 = m (\phi/\phi_0) + g$, where $N_0$ is the total number of sites of the square lattice and $\phi=Ba^2_0$.

   \item Note that the St\v{r}eda formula $\sigma_{xy} = e\left.\frac{\partial N(E)}{\partial B}\right|_{E=E_F}$ is in fact a fundamental thermodynamic relation\cite{P.Streda1982_IQHE2025cite15,WIDOM1982474,P.Streda1983}, irrespective of the specific details of the system,  
 where $N(E)=N_{occ}/(N_0 a^2_0)$ is electron density. This leads directly to the quantized Hall conductance $\sigma_{xy} = me^2/h$ for arbitrary magnetic fields.
   
\end{enumerate}
Crucially, our theory eliminates the fundamental constraint of rational magnetic flux  inherent in conventional IQHE frameworks, and establishes a unified description valid for arbitrary magnetic fields. More profoundly, in our theory, the quantum Hall conductance is fundamentally a quantization imposed by geometric constraints arising from Bragg planes, irrelevant to the conventional topological band (Chern number) picture. Therefore, it in fact provides a new paradigm for IQHE.



\emph{IEB spectrum of IQHE.}---Let us first illustrate how to calculate the energy spectrum of the square lattice in an arbitrary magnetic field with the IEB theory.  

 We consider a square lattice with $t_1$ and $t_2$ as the nearest neighbor hopping in $x$ and $y$ directions, respectively.  Applying a magnetic field and using the Landau gauge $A=(0,Bx,0)$, the Hamiltonian  is equivalent to an one dimensional atomic chain under a quasiperiodic potential, i.e., AAH model. 
\begin{equation}
    H(k_y) = t_1\sum_{j} \left( c_{j+1}^{\dagger} c_{j} + \mathrm{H.c.}\right) +\sum_{j} V_j c_{j}^{\dagger} c_{j},
\end{equation}
where $k_y$ is a good quantum number, $j$ is the site index along $x$ direction, and $V_j=2t_2\cos\left( k_y a_0 + G_2 \cdot R_j \right)$. Here, $R_j=j a_0$, $G_2=\alpha G_1$  is just the reciprocal lattice vector of the quasiperiodic potential, with $\alpha \equiv \phi / \phi_0$ and $G_1=2\pi/a_0$.  

Then, we can use the IEB theoy to solve $H(k_y)$.  
 Without loss of generality, we first assume $t_1>t_2$. In this case, the AAH model only have extended states, being localized in momentum space. Thus, we turn to momentum space, using  the Bloch waves of the atomic chain $|k_x\rangle$ as basis. 
 \begin{equation}{\label{AAH_k_space_BZ}}
    H(k_y)= \sum_{k_x}( \tau c^{\dagger}_{k_x+G_2}c_{k_x}+\mathrm{H.c.})+\sum_{k_x} \varepsilon_0(k_x) c^{\dagger}_{k_x}c_{k_x} 
\end{equation}
Here, $k_x \in \left[-G_1/2,G_1/2 \right]$ is the crystal momentum, $\tau \equiv t_2 \mathrm{e}^{ik_y a_0} $, and $\varepsilon_0(k_x)=2t_1\cos(k_x\cdot a_0)$. Eq.~\eqref{AAH_k_space_BZ} indicate that only the Bloch waves in the set $Q_{k_x} = \{ |K_m\rangle : K_m \equiv k_x + m G_2, m \in \mathbb{Z} \}$ are coupled. Note that, for an irrational magnetic flux $\alpha=G_2/G_1$, the coupled $|K_m \rangle$ form an infinite set, and the crystal momentum $K_m$ will densely fill the entire BZ $\left[-G_1/2, G_1/2 \right]$, called primary BZ (PBZ). In contrast, with a rational $\alpha$, $Q_{k_x}$ is a finite set. There features are reflected more clearly in their matrix forms. When $\alpha$ is irrational, $H(k_y)$ becomes an  infinite-dimensional matrix
\begin{equation}\label{Hmatrix-irrational}
    H(k_y) = 
    \begin{bmatrix}
        \ddots& &\cdots& & 0\\
          &\varepsilon_{-1}&\tau& & \\
        \vdots& \tau^{\ast}&\varepsilon_{0}&\tau&\vdots\\
         &  &\tau^{\ast}&\varepsilon_{1}& \\
         0&  & \cdots& &\ddots
    \end{bmatrix},
\end{equation}
where $\varepsilon_{m} = 2t_1\cos(K_m\cdot a_0)$ is the eigenenergy of $|K_m\rangle$. Meanwhile, for a rational $\alpha=p/q$, we get a finite-dimensional matrix under periodic boundary condition. 
\begin{equation}
    H(k_y) = 
    \begin{bmatrix}
        \varepsilon_{0}&\tau& 0&\cdots&\tau^{\ast}\\
        \tau^{\ast}& \varepsilon_{1}&\tau&\cdots&0\\
        0&\tau^{\ast}&\varepsilon_{2}&\cdots&0\\
        \vdots&\vdots&\vdots&\ddots&\vdots\\
         \tau&0&0&\cdots&\varepsilon_{q-1}
    \end{bmatrix}.
\end{equation}

We now elucidate the concept of IEBs. In the framework of IEB, only the basis $|K_0\rangle$ ($\varepsilon_0$ in the matrix) corresponds to the physical system, while all other $|K_m\rangle$  $(m \neq 0)$ represent replica bands--- artificially induced redundant degrees of freedoms---to account for the coupling. This picture is evident when we set the coupling $\tau=0$, since $|K_0\rangle$ and $\varepsilon_0$ are just the eigenstates and eigenvalues of the atomic chain then.  It is analogous to the Nambu
representation formalism in BCS superconductivity theory. A key feature is that, since  the eigenstates are localized states in k-space,  the eigenstate of the  physical system should be localized around the basis $|K_0\rangle$, while all redundant eigenstates are correspondingly localized around their respective redundant basis. Therefore, for a given $k_x$, we identify one real eigenstate here, and others are replica ones. Then, scanning $k_x$ across entire PBZ yields the IEB, an energy-band-like energy spectrum  for the quasiperiodic system. Here, although $k_x$ is not a good quantum number, it can still be the index of eigenstates. Note that, if $\alpha$ is rational, the IEB recovers to the conventional energy band with extended zone scheme. 

\begin{figure}[htbp]
	\centering
	\includegraphics[width=\linewidth]{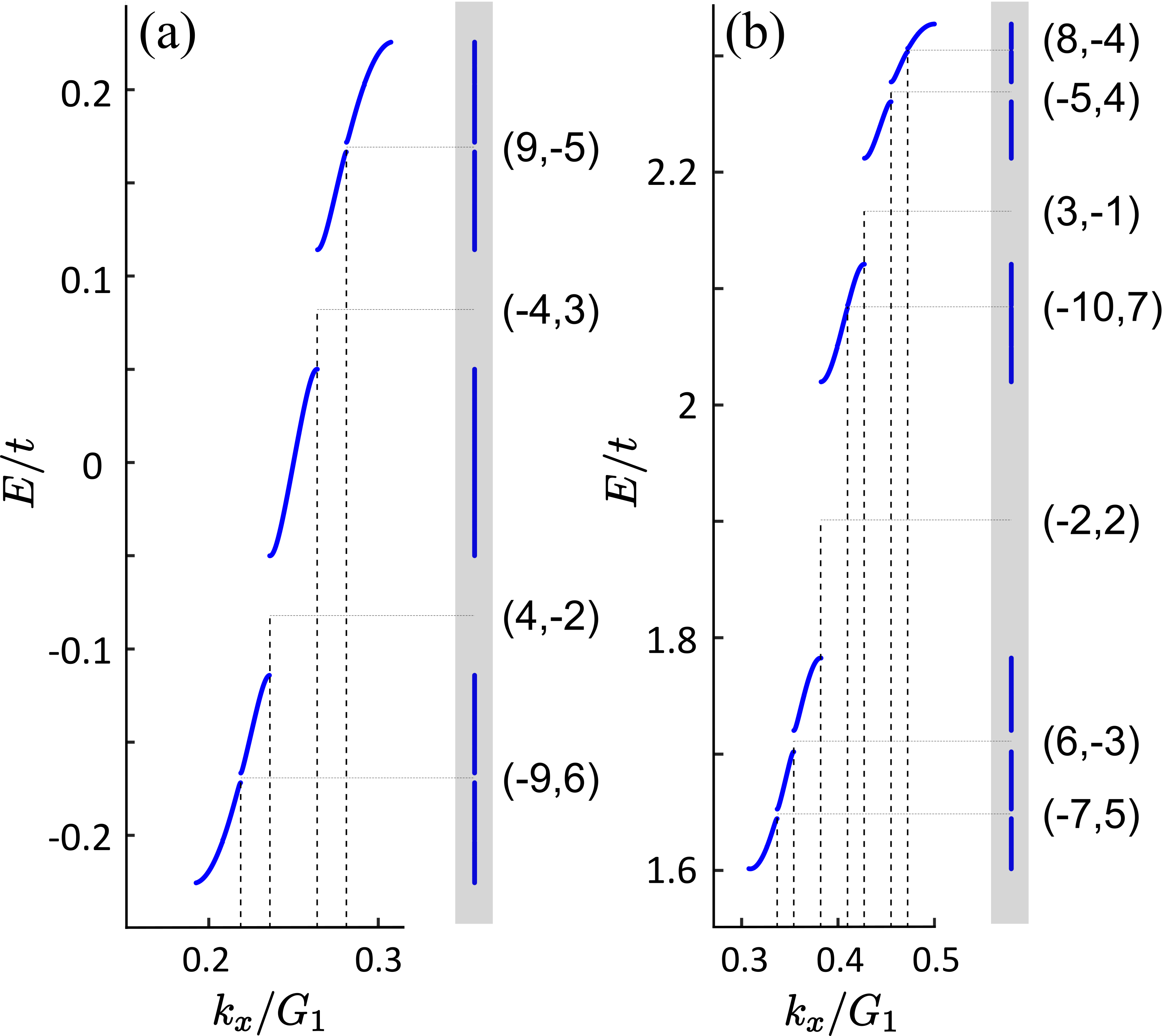}
    \caption{Enlarged IEBs of $\alpha = (\sqrt{5}-1)/2$ in different energy regions. Shadow region is the corresponding Hofstadter spectrum line. Vertical dashed lines represent the Bragg planes associated with the gaps. Gap labels $(m,g)$ are given as well. All the parameters are the same as Fig.~\ref{Hofstadter1}. }
    \label{IEBenlarged}
\end{figure}

In numerical calculations, we need to impose a cutoff $|m| < n_c$ on the infinite matrix in Eq.~\eqref{Hmatrix-irrational}. Then, for a fixed $k_y$, we diagonalize the Hamiltonian matrix to obtain all eigenstates, and  identify the one belonging to the IEB by examining their wave function profile in k-space.  Since all these eigenstates are localized in k-space,  a small truncation $n_c$ achieves sufficient accuracy.  A critical feature for a irrational magnetic field is that the eigenvalues of matrix in Eq.~\eqref{Hmatrix-irrational} are independent of $k_y$, in contrast to the rational cases. Therefore, all allowed $k_y$ yields identical IEBs. The calculation procedures for a rational magnetic field are similar, but exhibiting two key differences: the Hamiltonian matrix is finite, and the eigenstates are $k_y$-dependent. So,  no truncation is required then, and $k_y$ need to be scanned across the entire BZ.

The numerical results are shown in Fig.~\ref{Hofstadter1}~\footnote{See Supplemental Material at [URL] for the THE HOFSTADTER BUTTERFLY WITH ZERO AND EQUAL HOPPING STRENTH}. We set $t_1=1$ and $t_2=0.75$. For the typical irrational magnetic field $\alpha=\frac{\sqrt{5}-1}{2}$, the IEB is plotted in Fig.\ref{Hofstadter1} (c), see the blue solid lines. Other solid lines here denote the replica bands, where the orange (azure) lines are the replica band corresponding to $|K_1\rangle$ ($|K_{-1}\rangle$). Since IEB of an irrational $\alpha$ is independent of $k_y$, the IEB in Fig.~\ref{Hofstadter1} (c) exactly corresponds to a line in the Hofstadter spectrum, i.e. the blue lines in Fig.~\ref{Hofstadter1} (b). In contrast, for a rational $\alpha$, due to the magnetic translation symmetry,  IEBs revert to conventional energy bands ,  where $(k_x, k_y)$ is defined on the region $\left[-G_1/2,G_1/2 \right] \times \left[-G_1/2,G_1/2 \right]$ with extended zone scheme along $k_x$ direction. To illustrate the width of the energy bands, we calculated the bands  along the direction $k_x = k_y$, as indicated by the red lines in Fig.~\ref{Hofstadter1} (a) with $\alpha = \frac{3}{7}$. We plot the replica bands as well, where the orange (azure) lines correspond to basis $|K_1\rangle$ ($|K_{-1}\rangle$). Similarly, once all the momentum information is neglected, the energy bands exhibit  exact mapping to the Hofstadter spectrum at  $\alpha = \frac{3}{7}$, see red lines in Fig.~\ref{Hofstadter1} (b). So, the IEB theory provides a unified approach to calculate the Hofstadter spectrum no matter whether $\alpha$ is rational or not. We present the Hofstadter spectrum calculated via IEB method in Fig.~\ref{Hofstadter1} (b).

The IEB theory reveals the momentum-space distribution of eigenstates of a square lattice under a magnetic field---information missing from the Hofstadter spectrum but directly encoded in its spectral function, i.e., IEB structure.
Note that the spectral function of a AAH model\cite{XinyuGuo2024,JinrongChen2025} can be directly  measured in artificial platforms, e.g., the cold-atom systems\cite{Roati2008anderson}. We can expect a continuous variation of the spectral function regardless of whether $\alpha$ is rational or irrational, as illustrated in Fig.~\ref{Hofstadter1}, thereby providing a direct verification of our theory.

\begin{figure*}[tbp]
	\centering
	\includegraphics[width=0.85\textwidth]{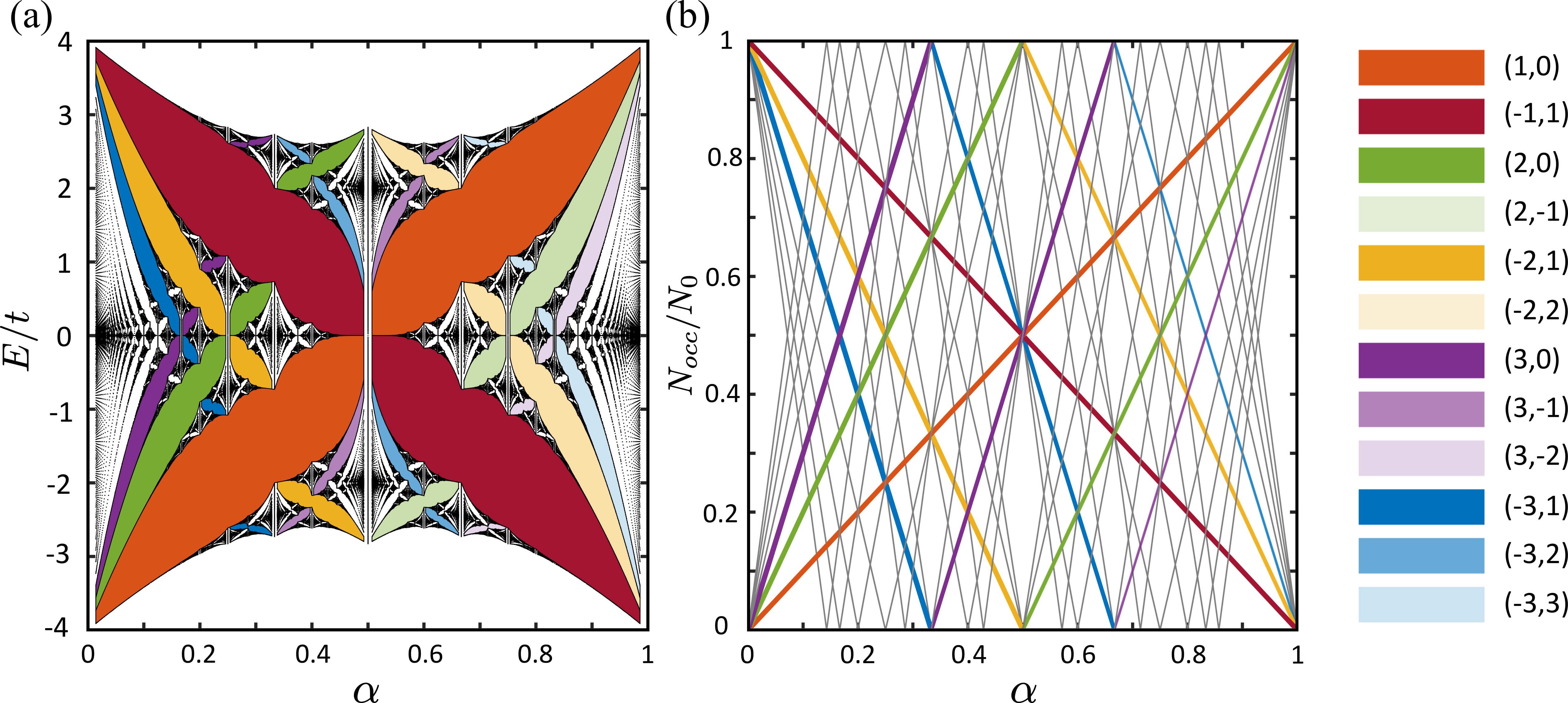}
    \caption{(a) is the Hofstadter spectrum with $t_2 = t_1$. Several largest enclosed domains are colored according to their gap labels $(m,g)$.   (b) is the corresponding Wannier diagram.}
    \label{Wannier}
\end{figure*}

\emph{Gap labeling rules.}---As shown in Fig.~\ref{Hofstadter1}, the energy gaps in the Hofstadter butterfly spectrum have an exact correspondence with those in the IEB spectrum. Within IEB framework, the gap in IEB actually arises from the energy degeneracy between the basis $|K_0\rangle$ and $|K_m\rangle$. Consequently, the gap position in k-space is determined by the condition  $\varepsilon_0(k_x)=\varepsilon_m(k_x)$, which gives ~\footnote{See Supplemental Material at [URL] for the DERIVATION OF THE POSITION OF BRAGG PLANES.} 
\begin{equation}
    k_x = \frac{mG_2 + gG_1}{2}, \quad 
    \text{with} \quad
    \begin{cases} 
        k_x \in \mathrm{PBZ} \\ 
        m, g \in \mathbb{Z}
    \end{cases}
    \label{Braggk}
\end{equation}
These $k_x$ are just the Bragg planes, which are represented as the vertical dashed lines in Fig.~\ref{Hofstadter1} (a) and (c). So, each Bragg plane is naturally and uniquely  described by two integers $(m,g)$, where the integer $m$ originates from the replica band 
$\varepsilon_m$ that generates the corresponding Bragg plane, and $g$ is an integer that determine  how  the momentum $mG_2$ is fold back to the region $\left[-G_1,G_1\right]$ by translation  $gG_1$.  Note that for a fixed $m$,  the requirement $k_x \in \mathrm{PBZ}$ leads two possible value of $g$, i.e., two distinct Bragg planes,  where $g=\lceil -m\alpha \rceil$ and $g=\lfloor -m \alpha \rfloor$ representing the Bragg planes  at positive and negative $k_x$, respectively. In Fig.~\ref{Hofstadter1} (c), 
the Bragg planes $(m,g)=(1,0)$ and $(m,g)=(1,-1)$ correspond to the two gaps associated with the replica band $\varepsilon_{1}$. And the gaps from $\varepsilon_{-1}$ are labeled as $(-1,0)$ and $(-1,1)$.

We then define the gap label.  As shown in Fig.~\ref{Hofstadter1} (c), an energy gap always relate to  two Bragg planes, $(m,g)$ and $(-m,-g)$.  Without loss of generality, we use index $(m,g)$ of the Bragg plane at $k_x>0$ to  label the gap, where $m$ originates from the related replica band $\varepsilon_m$ and $g=\lceil -m\alpha \rceil$. It thus provide a gap labeling rule for the IQHE.  
Such gap labeling rule is also valid for the rational magnetic field.  For a rational $\alpha=p/q$, we need  an additional constraint $|m|< q/2$ to prevent multivaluedness. In fact, we note that for the AAH model, the gap with a given $m$ varies continuously with $\alpha$. Thus, in practical calculations, although the corresponding Bragg planes are not unique when $\alpha$ is a rational magnetic flux, the gap label can still be uniquely determined using the results for irrational $\alpha$.

The gap labeling rule can be directly applied to the fractal Hofstadter butterfly spectrum. IEB for a irrational $\alpha$ actually has infinite energy gaps, which exhibit an exact corresponds to fractal spectrum in Hofstadter butterfly.  It is because the $k_x$ of Bragg planes never coincide with each other, and densely fill the PBZ. Therefore, the gaps arising from the Bragg planes are infinite. We plot the enlarged IEB in two distinct energy regions in Fig.~\ref{IEBenlarged}. More gaps become visible now, where the corresponding Bragg planes (vertical dashed lines), Hofstadter spectrum lines (shadow region), and gap labels of the Hofstadter spectrum are given as well.  

In addition,  the gap labels $(m,g)$ is irrelevant to the special values of $t_1$ and $t_2$, since they do not influence the the position of  Bragg plane as shown in Eq.~\eqref{Braggk}. In other words, adjusting $t_1$ and $t_2$ alters the magnitude of the energy gap, yet preserves both the gap position in k-space and the sequential order of eigenstates. Consequently, hopping exerts no effect on the gap label.

\emph{Electron density relation.}---
The IEB structure clearly illustrates a critical fact that, when $E_F$ is in a gap, the total number of occupied electron $N_{occ}$ is proportional to the k-space volume $\Delta k_x$ between the two Bragg planes associated to this gap. Note that the  labels of these two Bragg planes are $(m,g)$  and $(-m,-g)$. Using Eq.~\eqref{Braggk}, we obtain an electron density relation
\begin{equation}\label{densityrelation}
   \frac{N_{occ}}{N_0}=\frac{\Delta k_x}{G_1} = m\alpha + g,
\end{equation}
where $(m,g)$ are the gap labels, $N_0$ is the total nubmer of the lattice sites. 
This relation holds for both rational and irrational $\alpha$. Note that the density relation Eq.~\eqref{densityrelation}  differs fundamentally  from the Diophantine equation. In Eq.~\eqref{densityrelation}, both $m$ and $g$ are known quantities, whereas in the Diophantine equation within the TKNN theory, these two integers are unknown ones to be solved for.

\emph{Quantized Hall conductance.}---Note that the celebrated St\v{r}eda formula $\sigma_{xy} = e\left.\frac{\partial N(E)}{\partial B}\right|_{E=E_F}$ is in fact a general fundamental thermodynamic relation, irrespective of the specific details of the system.  Here, $N(E)=N_{occ}/(N_0 a^2_0)$ is electron density. Therefore, combining electron density relation and the St\v{r}eda formula, if $E_f$ lies in a gap, we finally obtain the quantized Hall conductance formula, 
\begin{equation}
    \sigma_{xy} = me^2/h
\end{equation}
where $m$ is the gap label, instead of the conventional Chern number. This is just the theory of IQHE in the framework of IEB picture, which is the central results of our work and holds for both rational and irrational magnetic fields. 

\emph{Wannier diagram.}---The IEB-based IQHE theory provides systematic theoretical elucidation of the Hofstadter spectrum and Wannier diagram. 

In Fig.~\ref{Wannier} (a), we first plot the standard Hofstadter spectrum of the square lattice with $t_1=t_2$ by the IEB method.  As mentioned above, for a given $\alpha$, a gap in Hofstadter spectrum  originates from the gap in IEB, with the gap label $(m,g)$. As $\alpha$ varies continuously, the energy gaps of IEBs undergo continuous evolution, thereby giving rise to enclosed  domains within the Hofstadter butterfly spectrum. Each enclosed  domain corresponds to an energy gap formed by the same replicated band $\varepsilon_m$, and is therefore uniquely indexed by the gap label $(m,g)$.  In Fig.~\ref{Wannier} (a), we color several largest domains according to their gap labels. Here, the index $m$ also represents the Hall conductance in the gap. Interestingly, the relation  $g=\lceil -m \alpha \rceil$ and $\alpha \in (0,1)$ indicates that the same $m$ in Hofstadter spectrum can provide different $g$. For example, when $m=1$, $g$ has only one possible value, $g = 0$; whereas when $m=2$, $g = -1$ when $\alpha \in (\frac{1}{2},1)$ and $g = 0$ when $\alpha \in (0,\frac{1}{2})$.
Based on the discussions above, we see that the IQHE theory presented here provides a comprehensive explanation of the principal characteristics of the Hofstadter butterfly spectrum.

The IEB theory provides a very natural explanation about the Wannier diagram. In Fig.~\ref{Wannier} (a), the orange domains correspond to the same gap $(1,0)$ with $\alpha \in (0,1)$. According to the electron density relation, this gap corresponds to a straight line defined by $(m,g)=(1,0)$ in the $N-\alpha$ diagram on the region $\alpha \in (0,1)$, see in Fig.~\ref{Wannier} (b). Similarly, the light green domain then corresponds to a line define by $(m,g)=(2,-1)$ on the region  $\alpha \in (\frac{1}{2},1)$. Therefore, the electron density relation actually give rise to the famous Wannier diagram\cite{Wannier1978}, where each gap in Hofstadter spectrum is represented by a straight line defined by the gap label $(m,g)$. 

\emph{Summary.}---In summary, we have established an unified IQHE theory for arbitrary magnetic fields, both rational and irrational, thereby transcending the conventional IQHE paradigms.  
Notably, the quantized Hall conductance here  is fundamentally interpreted as a quantization imposed by geometric constraints arising from Bragg planes, independent of the Chern number concept.  Thus, the quantized Hall conductance now  has two completely different  theoretical frameworks.  It yields a relation $ C=\Delta m$, where $C$ is the Chern number of a magnetic subband for a rational $\alpha$, and $\Delta m$ is the  difference in $m$-values between the two energy gaps adjacent to the band. 

At last, we think that our theory applies to any lattice model that can be mapped to a 1D quasiperiodic problem under an appropriate gauge choice.

\begin{acknowledgments}
    This work was supported by  the National Key Research and Development Program of China (Grants No.~2022YFA1403501), the National Natural Science Foundation of China (Grants No.12474169).
\end{acknowledgments}

\bibliography{reference}

\clearpage
\onecolumngrid
\setcounter{section}{0}
\renewcommand{\thesection}{S\arabic{section}}
\setcounter{secnumdepth}{1}

\begin{center}
\textbf{\large Supplementary Materials for: Theory of Integer Quantum Hall Effect in Irrational Magnetic Field}

\vspace{0.4cm}

{\normalsize
Zhao-Wen Miao$^{1}$,
Chen Zhao$^{1}$,
Jin-Hua Gao$^{1,*}$,
and X. C. Xie$^{2,3,4}$
}

\vspace{0.25cm}

{\small
\textit{$^{1}$School of Physics and Wuhan National High Magnetic Field Center,\\ Huazhong University of Science and Technology, Wuhan, Hubei 430074, China\\
$^{2}$Interdisciplinary Center for Theoretical Physics and Information Sciences, Fudan University, Shanghai 200433, China\\
$^{3}$International Center for Quantum Materials, School of Physics, Peking University, Beijing 100871, China\\
$^{4}$Hefei National Laboratory, Hefei 230088, China\\[0.2cm]
}}

\vspace{0.6cm}
\end{center}

\setcounter{equation}{0}
\setcounter{figure}{0}
\setcounter{table}{0}
\setcounter{page}{1}
\makeatletter
\renewcommand{\theequation}{S\arabic{equation}}
\renewcommand{\thefigure}{S\arabic{figure}}
\renewcommand{\bibnumfmt}[1]{[S#1]}

\section{Derivation of the position of Bragg planes}
    Within IEB framework\cite{ZheHe2024,XinyuGuo2024,JinrongChen2025}, the gap in IEB actually arises from the energy degeneracy between the basis $|K_0\rangle$ and $|K_m\rangle$. Consequently, the gap position in k-space is determined by the condition  $\varepsilon_0(k_x)=\varepsilon_m(k_x)$, where $\varepsilon_m(k_x) = 2t_1\cos(k_xa_0 + mG_2a_0)$, which gives 
    \begin{equation}
        \cos (k_x a_0) = \cos (k_x a_0 + mG_2 a_0)
    \end{equation}
    we get
    \begin{equation}
        F(k_x) = \cos (k_x a_0 + mG_2 a_0) - \cos (k_x a_0) = -2\sin \left(\frac{2k_x + mG_2}{2}a_0\right) \sin \left( \frac{mG_2}{2}a_0 \right) = 0
    \end{equation}
    Solving $F(k_x) = 0$ we get
    \begin{equation}
        \frac{2k_x + mG_2}{2}a_0 = \pi g,\quad m,g\in \mathbb{Z}
    \end{equation}
    without loss of generality we replace $-m$ by $m$, which finally leads to
    \begin{equation}
        k_x = \frac{mG_2 + gG_1}{2},\quad m,g\in \mathbb{Z}
    \end{equation}
    Considering the distance between any two Bragg planes
    \begin{equation}
        \Delta k_x = k_{x} - k_{x}' = \frac{m-m'}{2}G_2 + \frac{g-g'}{2}G_1,\quad m,m',g,g'\in \mathbb{Z} 
    \end{equation}
    For the specific condition $m = -m'$ and $g = -g'$, this leads to Eq.(6) in the main text
    \begin{equation}
       \frac{N_{occ}}{N_0}=\frac{\Delta k_x}{G_1} = m\alpha + g,
    \end{equation}
    
\section{The Hofstadter butterfly with zero and equal hopping strength}
    
    \begin{figure*}[tbp]
    	\centering
    	\includegraphics[width=0.9\textwidth]{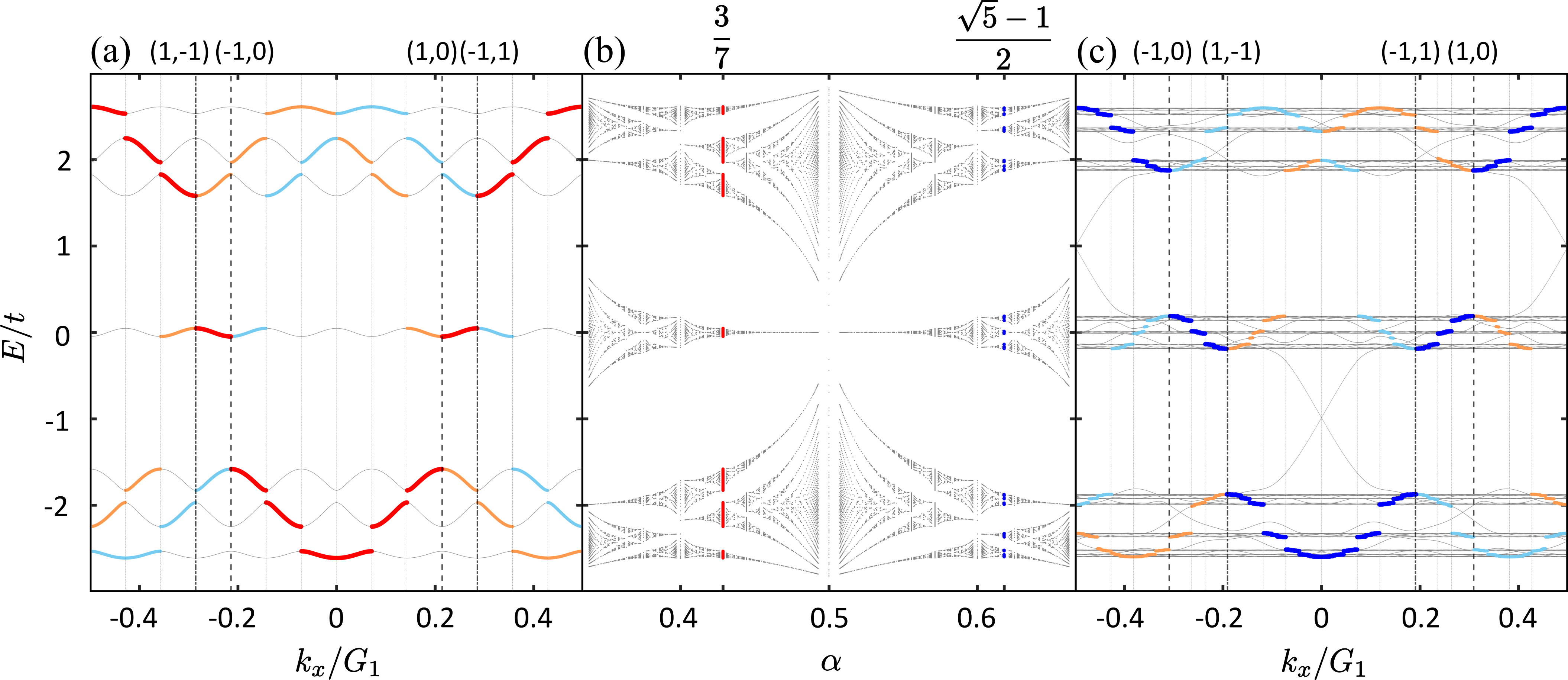}
         \caption{ IEBs and Hofstadter spectrum of a square lattice with  $t_2 = t_1$. (a) is the IEB of $\alpha = 3/7$ (red lines, rational case) along $k_y=k_x$ direction in BZ.  (b) is the Hofstadter spectrum in selected energy region. (c) is the IEB of $\alpha = (\sqrt{5}-1)/2$  (blue lines, irrational case). In (a) and (c), the replica bands corresponding to $|K_1\rangle$ and $|K_{-1}\rangle$ are highlighted with orange and azure lines, respectively. The vertical dashed lines represent the Bragg planes. Here, the cutoff $n_c = 44$ is used, and  $t_1=t$ as energy unit. }\label{supplementary figure1}
    \end{figure*}

    \begin{figure*}[tbp]
    	\centering
    	\includegraphics[width=0.9\textwidth]{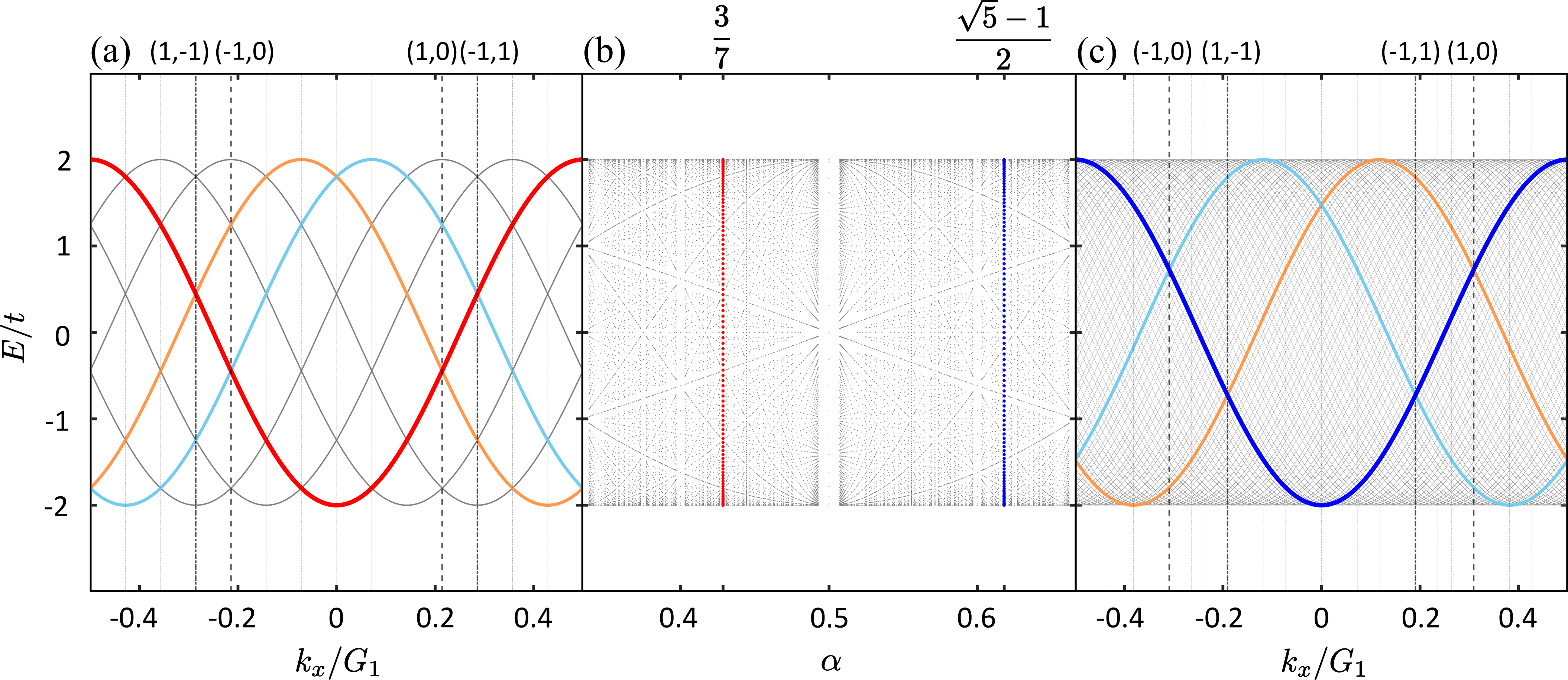}
         \caption{ IEBs and Hofstadter spectrum of a square lattice with  $t_2 = 0$. (a) is the IEB of $\alpha = 3/7$ (red lines, rational case) along $k_y=k_x$ direction in BZ.  (b) is the Hofstadter spectrum in selected energy region. (c) is the IEB of $\alpha = (\sqrt{5}-1)/2$  (blue lines, irrational case). In (a) and (c), the replica bands corresponding to $|K_1\rangle$ and $|K_{-1}\rangle$ are highlighted with orange and azure lines, respectively. The vertical dashed lines represent the Bragg planes. Here, the cutoff $n_c = 44$ is used, and  $t_1=t$ as energy unit. }\label{supplementary figure2}
    \end{figure*}
    
    Fig.~\ref{supplementary figure1} shows the Hofstadter butterfly spectrum and the IEBs with equal hopping strength , while Fig.~\ref{supplementary figure2} presents the results with $t_2$ turned off. It can be seen that changing $t_2$ does not alter the positions of the Bragg planes and turning on $t_2$ will open a gap at the degenerate points.

    It can be concluded that the Bragg planes are initially determined by the integers $m$ and $g$. Subsequently, the $IEB$ is selected via the incommensurate method. By combining these two steps, the energy gaps can be unambiguously identified and indexed. Importantly, in the incommensurate system, both $m$ and $g$ possess well-defined physical interpretations -- $m$ serves as an index of the $IEB$ and $g$ denotes the number of times the momentum $k_x + mG_2$ must be folded back into the $PBZ$ during the coupling process. All quantities are predetermined, with no undetermined parameters in the formulation.

\section{Further Discussion on Incommensurate Hamiltonian}
    The general form of an $n$-dimensional tri-diagonal matrix is as follows
    \begin{equation}
        A_n = 
        \begin{bmatrix}
            a_1&b_1& & &\cdots& \\
            c_1&a_2&b_2& & \cdots& \\
             &c_2&a_3&b_3& & \\
            \vdots& & & & \ddots& \vdots\\
              & & &\cdots&a_{n-1}&b_{n-1} \\
              & & &\cdots&c_{n-1}&a_n
        \end{bmatrix}
    \end{equation}
    with the determinant
    \begin{equation}
        \det(A_n) = a_n\det(A_{n-1}) - c_{n-1}b_{n-1}\det(A_{n-2})
    \end{equation}
    For the irrational Hamiltonian matrix
    \begin{equation}
        H = 
        \begin{bmatrix}
            \ddots& &\cdots& & 0\\
              &\varepsilon_{-1}&\tau& & \\
            \vdots& \tau^{\ast}&\varepsilon_{0}&\tau&\vdots\\
             &  &\tau^{\ast}&\varepsilon_{1}& \\
             0&  & \cdots& &\ddots
        \end{bmatrix},
    \end{equation}
    the eigenvalues are calculated by $\det(H-IE)$. The phase factor $\mathrm{e}^{ik_y a_0}$ in $\tau$ cancels out when calculating $\tau^{\ast}\tau$, so that the eigenvalues are $k_y$-independent.

\end{document}